\documentstyle[aps,prb,multicol,epsf]{revtex}

\def\Tr{\mbox{Tr}}

\newcommand{\bleq}{\ifpreprintsty
                   \else
                   \end{multicols}\vspace*{-3.5ex}{\tiny
                   \noindent\begin{tabular}[t]{c|}
                   \parbox{0.493\hsize}{~} \\ \hline \end{tabular}}
                   \fi}
\newcommand{\eleq}{\ifpreprintsty
                   \else
                   {\tiny\hspace*{\fill}\begin{tabular}[t]{|c}\hline
                    \parbox{0.49\hsize}{~} \\
                    \end{tabular}}\vspace*{-2.5ex}\begin{multicols}{2}
                    \fi}
\newcommand{\bcols}{\ifpreprintsty\else\begin{multicols}{2}\fi}
\newcommand{\ecols}{\ifpreprintsty\else\end{multicols}\fi}

 \newcommand {\be} {\begin{equation}}
\newcommand {\bea} {\begin{eqnarray} \nonumber }
\newcommand {\ee} {\end{equation}}
\newcommand {\eea} {\end{eqnarray}}
 \newcommand {\eps} {\epsilon}
 \newcommand {\psib} {\overline{\psi}}
 \newcommand {\nn} {\nonumber}
\def\(({\left(}
\def\)){\right)}
\def\[[{\left[}
\def\]]{\right]}

\begin{document}
\bibliographystyle{prsty}
\title{Wigner-Dyson  Statistics from the Replica Method}
  
\draft

\author{Alex Kamenev$^{1,2}$ and Marc M\'ezard$^{1,3}$ 
  }
\address{$^1$Institute for Theoretical Physics, 
 University of California Santa Barbara, CA 93106-4030, USA.\\
 $^2$Department of Physics, 
 University of California Santa Barbara, CA 93106-4030, USA. \\
 $^3$CNRS, Laboratoire de Physique Th\'eorique de l'ENS,  France.
  \\
  {}~{\rm (\today)}~
  \medskip \\
  \parbox{14cm} 
    {\rm We compute the correlation functions of the eigenvalues in the 
    Gaussian unitary ensemble using the fermionic replica method. We show
    that non--trivial saddle points, which break replica symmetry,  must be 
    included in the calculation in order to reproduce correctly the exact 
results
    for the  correlation functions at large distance. 
     }\bigskip \\ }

\maketitle


\section{Introduction}
\label{s1}

 Random Matrix Theory (RMT) has found  broad applications 
in physics ranging from nuclear spectra to electrons in metallic 
grains, see Refs. \cite{Mehta,Weidenmuller98} for  reviews. 
The first mathematically rigorous results for the level statistics were 
derived by Gaudin \cite{Gaudin} and Dyson \cite{Dyson} using the method 
of orthogonal 
polynomials. The progress in the theory of Anderson localization in 
the late seventies 
established a close relation between the RMT and 
matrix $\sigma$--models initially in their replicated form 
\cite{Wegner79,Efetov81}. Despite of a lot of efforts 
\cite{Khmelnitskii,Zirnbauer85}, however, the 
initial attempts to reproduce Dyson's results from the replicated 
$\sigma$--models were not successful: while the density of eigenvalues
could be found easily, it seemed that the eigenvalue correlations
could not be obtained from the replicated $\sigma$--models. 
The most detailed account of such attempts was probably given by 
Zirnbauer and Verbaarschot \cite{Zirnbauer85}, who computed both bosonic 
and fermionic replicated  $\sigma$--models and obtained different results, 
both  differing from the correct one. 
Only the supersymmetric (SUSY) 
formulation of the $\sigma$--model introduced by Efetov \cite{Efetov83} 
gave a correct-and beautiful- way of calculating  the correlations
of eigenvalues from a $\sigma$ model formulation.  
It has thus become a common wisdom for fifteen years that the SUSY is 
the only field theoretic method capable to compute  the RMT level statistics,
 while the internal 
subtleties of the replica method seem to make it inapplicable for this task. 

Looking at the replicated sigma model approach, 
it is clear that there is one
underlying assumption in the existing computations, which is the absence of any
spontaneous breaking of replica symmetry. The study of the statistics of
eigenvalues of a large $N\times N$ random matrix
is mapped exactly onto a $\sigma$--model, where the action is of order $N$.
Among  various saddle points of the
$\sigma$--model which could a-priori contribute, only the trivial 'replica 
symmetric'
 one was discussed. In this paper we revisit the problem, and  consider all 
possible
 saddle points. Because of the symmetry of the $\sigma$--model the saddle points
 are actually saddle point manifolds. We show that the computation of the 
 $r$--point correlation function involves $2^r$ 
 saddle point manifolds.  For the one--point function (the
 level density), the trivial saddle point gives Wigner's semi-circle law,
 whereas the second one contributes to order $1/N$, and is, in fact, needed  to 
obtain 
 the oscillatory component of the density of states (DOS).
 For the two--point function, the effect is even more dramatic since one of the
 extra three non--trivial saddle point manifolds  contributes to the leading 
order
 at large $N$ (the other two being $1/N$ corrections). Taking it into
 account gives the correct result for the two--point correlation function.
  A similar situation was actually found in the SUSY, where
Andreev and Altshuler \cite{Andreev95} showed that the asymptotic
behavior of the two level correlations at large energy differences
 (in  units of the mean level spacing) can be obtained by the saddle point 
evaluation of the SUSY $\sigma$--model, including one extra
non--trivial saddle point.

We restrict ourselves here to the Gaussian unitary ensemble (GUE), 
where we derive  the exact results 
for the DOS, including oscillatory $1/N$ correction
 as well as the large energy difference behavior of the two--point correlation 
function
 using the fermionic replica formalism.
 The strategy is simple enough to be easily 
generalized for the higher correlation functions for which it also leads to 
known GUE expressions. Technically, a very useful step is to integrate exactly 
over
the angular  degrees of freedom of the $\sigma$--model, using
the Itzykson--Zuber integral \cite{Itzykson80}, which leaves
 the $n$ eigenvalues of
the replicated $\sigma$--model as the only integration variables. The saddle
points are thus discussed on the level of eigenvalues, and the resulting 
replica symmetry breaking appears to be a particularly simple version of 
the vector replica symmetry breaking mechanism encountered in several
disordered systems \cite{dotsenko}.

The paper is organized as follows: in section \ref{s2} we present the 
calculations of the average DOS, including the $1/N$ corrections. 
This section is also used to introduce 
notations and illustrate the technique on this simple, but instructive 
example. Calculation of the two--point correlation function is given 
in section \ref{s3}. Finally sections \ref{discussion} and \ref{s5} are devoted 
to discussions of the method and remaining open questions. 
Some technical details are presented in three appendices.

\section{Density of States} 
\label{s2}

We are interested in the spectral properties of the 
random $N \times N$ Hermitian Hamiltonians, $H$, with the Gaussian 
probability distribution function 
\be
P(H) =  2^{N(N-1)/2} \left( \frac{N}{2\pi} \right)^{N^2/2} 
\exp\left\{ -{N \over 2} \Tr\, H^2 \right\} \, .
						\label{ensemble}
\ee
We begin with the calculation of the one--point  function $S_1(E)$
defined as 
\begin{equation} 
S_1(E) = N^{-1}
\overline{\Tr(E-H)^{-1} }\, ,
                                                              \label{q1}
\end{equation}  
where the complex energy $E$ has an infinitesimal negative imaginary part.
The bar stands for the averaging over $H$ with the measure given by 
Eq.~(\ref{ensemble}). 
The DOS is given by the imaginary part of this correlation 
function 
\begin{equation} 
\nu(E) = \pi^{-1} \Im S_1(E)\, .
                                                              \label{q1a}
\end{equation}  
Introducing $2 N$ anti-commuting variables $\psi_x, \psib_x$, 
where $x$ is a discrete index
$x \in \{ 1,\ldots N\}$, the  correlation function 
may be written as
\begin{equation} 
S_1(E) = N^{-1}
\overline{{\partial \over \partial E} \ln Z(E)  }\, , 
                                                              \label{q2}
\end{equation} 
where 
\begin{equation} 
Z(E) = \int \prod_{x=1}^N d \psib_x d \psi_x \ \exp\left\{
-\sum\limits_{x,y=1}^{N} 
\psib_x
\left( E \delta_{xy} - H_{xy}\right)
\psi_{y}\right\}   \, .
                                                              \label{q3}
\end{equation} 
We use now  the replica trick to write the logarithm  as 
$\ln Z = \lim_{n\to 0}(Z^n-1)/n$. 
As a result the  correlation function takes the form 
\begin{equation} 
S_1(E) = \lim_{n\to 0} {1\over n } S_1^{(n)}(E)\, ; \hskip 1cm
S_1^{(n)}(E) = N^{-1} {\partial \over \partial E}\overline{Z(E)^n}\, , 
                                                              \label{q5}
\end{equation} 
It is convenient to introduce  the 
 generating function $Z^{(n)}(\hat E) \equiv Z(E_1) \ldots Z(E_n)$, where $\hat 
E$ is a diagonal 
$n\times n$ matrix which has the form 
$\hat E =\mbox{diag}\{E_1,\ldots ,E_{n} \}$. In the limit $E_1,\ldots E_{n}\to 
E$ 
the generating function $Z^{(n)}(\hat E)$ approaches  $Z(E)^n$ and gives the 
one-point function according to  Eq.~(\ref{q5}). 
The generating function $Z^{(n)}(\hat E)$ may be  written 
as the fermionic integral analogous to that of Eq.~(\ref{q3}). The integration 
runs now over $2nN$ fermionic variables $\psi_{x}^j,\psib_{x}^j$, where 
$x\in \{1\ldots N\}$ and $j\in \{1\ldots n\}$. Performing the  averaging 
over the  random matrix $H$ one finds  
 \be
 \overline{ Z^{(n)}(\hat E)} = \int \prod_{x=1}^N \prod_{j=1}^n d \psib_x^j
 d \psi_x^j \ 
\exp\left\{ 
-\sum_{x,j} \psib_x^j E_j \psi_x^j -{1 \over 2 N} \sum_{j,k=1}^n
  T_{jk}T_{kj}
 \right\}
 \ ,
                                                              \label{zn}
 \ee
where $T_{jk}\equiv \sum_x \psib_x^j \psi_x^k$. Introducing the 
Hubbard--Stratonovich decoupling of the last term with the $x$--independent,  
Hermitian $n\times n$ matrix $\hat Q$ and integrating out the anti--commuting 
variables, one obtains the $\sigma$--model in the form 
\cite{Wegner79,Efetov81}:
\begin{equation} 
\overline{Z^{(n)}(\hat E) } = c_n
\int\!\!  d[\hat Q] \exp \left\{ 
-{N\over 2}\, \Tr\,  {\hat Q}^2 + N\, \Tr \ln(\hat E - i \hat Q) \right\}
\, , 
                                                              \label{q6}
\end{equation} 
\begin{equation} 
c_n= \left( \frac{N}{2\pi} \right)^{n^2/2} 2^{n(n-1)/2}\, . 
                                                              \label{q6a}
\end{equation}

The standard route to study the large $N$ limit of the DOS prescribes to 
take all energies equal to $E$, look for 
the saddle points of the functional integral, Eq.~(\ref{q6}), and then 
consider fluctuations around the saddle points (see e.g. \cite{Zirnbauer85}).
The saddle point equation, $\hat Q=(\hat Q+i E)^{-1}$, may be solved by going 
to the basis
where $\hat Q$ is diagonal. One finds two possible solutions for each
of the eigenvalues $\lambda_1,\ldots ,\lambda_n$
of $\hat Q+ iE$: 
\begin{equation} 
\lambda_{\pm}(E) = 
\pm \sqrt{1-{E^2\over 4} } + {iE\over 2} \, .
                                                              \label{dos1}
\end{equation} 
This results in $2^n$ distinct diagonal matrices. As $\hat Q$ is obtained from
the diagonal matrix through a unitary transformation, one finds $n+1$ saddle
point manifolds, generated by rotations of
${U(n)}/[{U(p) U(n-p)}]$, applied to a diagonal matrix with 
$\lambda_i= \lambda_{-}$ for $ i=1\ldots p$ and 
$\lambda_j= \lambda_{+}$ for $j=p+1\ldots n$, where $0\leq p \leq n$.   
Which of these manifolds dominates is not obvious a-priori
\cite{fn_sadpo}. It 
is known that the leading, large $N$, contribution to the DOS is given
by the simple `replica symmetric' saddle point 
(this one is just a point not a manifold, since $\hat Q$
is proportional to $\openone$)
with $p=0$  \cite{Zirnbauer85}. 
We show below   that other saddle point manifolds are crucial for 
computations of the $1/N$ corrections to the DOS, 
as well as the level correlations.
In order to take into account all possible saddle points, 
one must be careful to integrate 
properly over the manifolds. To achieve this we found it convenient to 
employ  the method introduced by T. Guhr \cite{Guhr91} in the SUSY context. 
This method takes advantage of the 
Itzykson--Zuber integral \cite{Itzykson80} for the GUE to integrate 
all rotational degrees of freedom exactly, and discusses the saddle points 
in terms of the eigenvalues of $\hat Q$. A similar method has been
employed by Br\'ezin and Hikami in order to derive and extend the universality
of level spacing distributions \cite{brezin}.

By shifting first the $\hat Q$ matrix 
 $\hat Q\to  \hat Q - i\hat E$ \cite{foot0} one obtains:
\be
\overline{Z^{(n)}(\hat E)} = c_n (-i)^{N n} 
\int d[\hat Q] \exp  \left\{ 
-{N\over 2} \Tr (\hat Q - i\hat E)^2 \right\} (\det  \hat Q)^N
 \label{shifted}
 \ee
The Hermitian matrix $\hat Q$ is then diagonalized by 
a unitary transformation,  $\hat Q= U^{-1} \hat \Lambda U$ with 
$U\in U(n)$ and the diagonal matrix 
$\hat \Lambda = \mbox{diag}\{\lambda_1,\ldots \lambda_{n}\}$. The volume 
element transforms like 
\begin{eqnarray} 
&&d[\hat Q] = \Delta_{n}^2(\hat \Lambda)\, d[\hat \Lambda]\, d\mu(U) \, ; 
                           \nonumber   \\
&&d[\hat \Lambda] = \prod \limits_{j=1}^{n} d\lambda_j \, , 
                                                              \label{q7}
\end{eqnarray} 
where $d\mu(U)$ is the measure of the group $U(n)$ and the Jacobian is 
given by the square of the Vandermonde determinant 
\begin{equation} 
\Delta_{n}(\hat \Lambda) \equiv  \prod \limits_{1\le i < j \le n}
(\lambda_j - \lambda_i) \, .
                                                              \label{q8}
\end{equation} 
The non--trivial integration over the group $U(n)$ in Eq. (\ref{shifted}) 
may be performed using the Itzykson--Zuber integral \cite{Itzykson80} 
(see also Appendix 5 of Ref. \cite{Mehta}): 
\bea
\overline{Z^{(n)}(\hat E)}&=& c_n(-i)^{N n} 
\int d[\hat \Lambda] \Delta_{n}^2(\hat \Lambda) (\det \hat \Lambda)^N
\int d\mu(U) \exp \left\{ 
-{N\over 2} \Tr (\hat \Lambda  - iU\hat E U^{-1})^2 \right\}   
                        \nonumber   \\
&=& c_n (-i)^{N n}  \left(\frac{\pi}{N}\right)^{n(n-1)/2} 
{1 \over \Delta_{n}(i\hat E)}
\int d[\hat \Lambda] 
{\Delta_{n}(\hat \Lambda)} 
(\det \hat \Lambda)^N 
 \exp \left\{ 
-{N\over 2} \sum\limits_{j=1}^{n} (\lambda_j  - iE_j)^2 \right\}\, .
                                                              \label{q9}
\end{eqnarray}  
It may seem that the last expression has poles at $E_i = E_j$, 
which is not the case. 
Indeed, the integral over the eigenvalues $\lambda_j$ results in a  
totally antisymmetric function of $E_i$, which vanishes if $E_i = E_j$. 
The reason  we  introduced the diagonal matrix 
$\hat E$ with all  elements different was to regularize properly this 
fictitious singularity. The next step is to take the limit 
$E_1,\ldots E_{n}\to E$. This 
procedure is described in  appendix \ref{app1}. The result is given by 
\begin{equation} 
\overline{Z^n(E) } = c_n'
\int \!\! d[\hat \Lambda] \Delta_{n}^2 (\hat \Lambda)
\exp\left\{
-N \sum\limits_{j=1}^{n} A(\lambda_j,E) \right\} \, ,
                                                              \label{q10a}
\end{equation}  
where 
\begin{equation} 
A(\lambda_j,E) = {1\over 2}(\lambda_j-iE)^2 - \ln \lambda_j\, ,
\hskip 1cm 
c_n'=(-i)^{Nn}\frac{N^{n^2/2} }{(2\pi)^{n/2} } 
\frac{1}{\prod\limits_{j=1}^{n} j!} \, .
                                                              \label{q11a}
\end{equation} 
Equation (\ref{q10a}) may be easily derived already from the first line of 
Eq.~(\ref{q9}) if $\hat E$ is proportional to the unit matrix. 
The reason the longer procedure was presented is to 
generalize it later for the case of  multi--point correlation functions. 
Employing 
Eq.~(\ref{q5}), one finally obtains for the replicated single--point 
correlation function
\begin{equation} 
S_1^{(n)}(E)  = ic_n'
\int \!\! d[\hat \Lambda] \Delta_{n}^2 (\hat \Lambda)
\exp\left\{
-N \sum\limits_{j=1}^{n} A(\lambda_j,E) \right\} 
\left[ \sum\limits_{j=1}^{n} (\lambda_j - iE) \right]\, . 
                                                              \label{q10}
\end{equation}

So far all the calculations were  exact. One may take now advantage 
of the large parameter $N\gg 1$ to perform the integrations in 
Eq.~(\ref{q10}) by the saddle point method. Differentiating $A(\lambda_j,E)$ 
over 
$\lambda_j$, one finds two saddle point solutions for each $\lambda_j$, 
which are given by $\lambda_{\pm}(E)$ defined by Eq.~(\ref{dos1}). 
This leads to $2^n$ distinct saddle points of the integral in Eq.~(\ref{q10}), 
each of them may be brought to the form 
\begin{equation} 
\hat\Lambda = \mbox{diag} \{
\underbrace{\lambda_-,\ldots \lambda_-}_{p}, 
\underbrace{\lambda_+,\ldots \lambda_+}_{n-p}  \} \, .
                                                              \label{q10b}
\end{equation} 
There are $C_n^p = n!/[p!(n-p)!]$ such saddle points for every $p$, $0\leq p 
\leq n$. 
Finding  dominant saddle points is not totally trivial. On the one
hand, the saddle point action, $A_{\pm}=A(\lambda_{\pm}(E),E)$, is such
that $|A_+|=|A_-|$ when the energy is real (see footnote \cite{fn_sadpo}). 
Furthermore,  
one must be careful with the saddle point calculation of the integral in  
Eq.~(\ref{q10}) because the preexponential factor, $\Delta_{n}^2 (\hat 
\Lambda)$, 
vanishes identically at any saddle point (for $n>2$). Therefore it should 
be expanded to sufficiently high power of $\lambda_j - \lambda_{\pm}$ to 
produce a non--zero result.  To this end we introduce variables $\xi_j$ 
describing fluctuations around the saddle point, Eq.~(\ref{q10b}), defined as: 
\bea
\lambda_i= &\lambda&_- +\xi_i/\sqrt{N}\, ;  \ \ \ \ \ i=1 \ldots p\, , 
                                                \nonumber   \\
\lambda_j= &\lambda&_+ +\xi_j/\sqrt{N}\, ;  \ \ \  \ \ j=p+1 \ldots n \, 
                                                          \label{q10c}
\eea
and the diagonal matrices $\hat \Xi_- =\mbox{diag}\{\xi_1\ldots 
\xi_p\}$
and $\hat \Xi_+ =\mbox{diag}\{\xi_{p+1}\ldots \xi_n\}$.
For any $p$ one  may identically rewrite  $\Delta_{n}^2 (\hat \Lambda)$  as
\begin{equation} 
\Delta_{n}^2 (\hat \Lambda)= \(({1 \over \sqrt{N} } \))^{p(p-1)+(n-p)(n-p-1)} 
\left[ 
\prod \limits_{i=1}^{p}\prod \limits_{j=p+1}^{n}
\left(
\lambda_+-\lambda_-+{\xi_j - \xi_i \over \sqrt{N}}
\right)
\right]^2
\Delta_{p}^2 (\hat \Xi_-) \Delta_{n-p}^2 (\hat \Xi_+) \, .
                                                              \label{q12}
\end{equation}  
The  factor in  square brackets on  the r.h.s. of 
this expression is non--vanishing at the saddle point 
and therefore may be substituted by its saddle point value. 
Expanding the exponent to  second order in the deviations 
from the saddle point, 
one obtains for the replicated correlation function 
\begin{eqnarray} 
S_1^{(n)}(E) = &&ic_n' \sum\limits_{p=0}^{n} C_n^p
\left[ \lambda_{+} - \lambda_{-} \right]^{2p(n-p)}
e^{- NpA_{-} - N(n-p)A_{+} } 
\Big[ p\lambda_{-} + (n-p)\lambda_{+} -inE \Big]
\(({1 \over \sqrt{N} } \))^{p^2+(n-p)^2}
 \times
                                          \nonumber                \\
&& \int\!  \prod \limits_{i=1}^{p} d\xi_i \Delta_{p}^2 (\hat \Xi_-)
\exp\left\{
-{1\over 2} A^{\prime\prime}_{-} \sum\limits_{i=1}^{p} \xi_i^2
\right\} \ 
\int \!\!\! \prod \limits_{j=p+1}^{n} d\xi_j \Delta_{n-p}^2 (\hat \Xi_+)
\exp\left\{
-{1\over 2} A^{\prime\prime}_{+}\!\!  \sum\limits_{j=p+1}^{n} \xi_j^2
\right\}  \, ,
                                                              \label{q13}
\end{eqnarray}  
where 
$A^{\prime\prime}_{\pm} = \partial^2_\lambda A(\lambda,E)|_{\lambda_{\pm} }$. 
The two remaining integrals are known 
as a version of Selberg's integral \cite{Mehta}, given by 
\begin{equation} 
\int \prod \limits_{i=1}^{p} d\xi_i \Delta_{p}^2 (\hat \Xi_-)
\exp\left\{
- t \sum\limits_{i=1}^{p} \xi_i^2
\right\} 
= 
(2t)^{-p^2/2 } \Omega_p \, ,\hskip .5cm 
\mbox{with} \ \ \ 
\Omega_p= (2\pi)^{p/2}
\prod\limits_{i=1}^{p} i!\,\, .
                                                              \label{q14}
\end{equation}  
As a result the correlation function  takes the form 
\begin{equation} 
S_1^{(n)}(E)  = (i)^{1-nN}  e^{-nNA_{+} }
\sum\limits_{p=0}^{n} F_n^p 
\left[ \lambda_{+}\! -\! \lambda_{-} \right]^{2p(n-p)}
\frac{N^{p(n-p)} } 
{(\sqrt{ A^{\prime\prime}_{+}})^{(n-p)^2}
(\sqrt{ A^{\prime\prime}_{-}})^{p^2}    }\,  
e^{pN(A_{+} - A_{-}) }
\Big[ p\lambda_{-}\! +\! (n-p)\lambda_{+}\! - \! inE \Big]   \, ,
                                                              \label{q15}
\end{equation}  
where we have introduced the $F_n^p$ symbol as 
\begin{equation} 
F_n^p \equiv  C_n^p \, 
\frac{\prod\limits_{j=1}^{p} j! \prod\limits_{j=1}^{n-p} j!}
{\prod\limits_{j=1}^{n} j!} = 
\prod\limits_{j=1}^{p} 
\frac{\Gamma(j)}{\Gamma(n-j+1)}\,; \hskip 1cm   p\neq 0\,  
                                                              \label{q16}
\end{equation}  
and $F_n^{p=0}=1$. Since the gamma function diverges at any negative integer,  
$F_n^{p>n}=0$. Therefore  one may extend the summation over $p$ in 
Eq.~(\ref{q15}) up to infinity. The resulting expression is suitable for the 
analytical continuation, $n\to 0$. To continue the $F_n^p$ function one may use 
the identity $\Gamma(-z)=-\pi/(\Gamma(z+1) \sin\pi z)$ with $z=j-1-n$, 
which leads to 
\begin{equation} 
F_n^p = (-1)^{ p(p-1)/2} 
\left[\frac{\sin \pi n}{\pi} \right]^p  
\prod\limits_{j=1}^{p} \Gamma(j) \Gamma(j-n) \,; 
\hskip 1cm   p\neq 0\,  .
                                                              \label{q17}
\end{equation}
Expanding the $\sin\pi n$ for small $n$, one obtains   
$F_{n\to 0}^{p=0}=1$;  $F_{n\to 0}^{p=1} = n$,
whereas $F_{n\to 0}^{p\geq 2}= O(n^p)$. As a result only two 
terms with $p=0$ and $p=1$ survive in the sum in Eq.~(\ref{q15}). 
This is an important conclusion: out of the $2^n$ possible saddle points 
only one with $p=0$ and $n$ with $p=1$ contribute in the replica limit.
Of course, the argument we have given here is somewhat heuristic since the
series in Eq.~(\ref{q15}) is divergent for non--integer $n$. 
This should be justified more rigorously. 
Although we have not completed the rigorous proof 
of this statement, we present some elements of the proof in appendix \ref{app2}.

One may easily check that the $p=1$ contribution is smaller by a factor 
$1/N$ with respect to the $p=0$ one.  We are therefore back to the 
familiar statement that the large $N$ limit of the DOS may be calculated 
by the replica method 
using a single trivial saddle point for the $\hat Q$ matrix which is 
proportional to the unit matrix. In particular one obtains 
\begin{equation} 
{S_1(E)}_{(p=0)} = i(\lambda_+ - iE)  \, .
                                                              \label{q18}
\end{equation}  
Employing Eqs.~(\ref{q1a}), (\ref{dos1}) one finds the famous ``law of 
semi-circle''
for the DOS, $\nu_0(E) = \sqrt{4 - E^2}/(2\pi)$. 

It is instructive to look at the $1/N$ contribution originating from the $p=1$ 
saddle point manifold:
\begin{equation} 
{S_1(E)}_{(p=1)} = - i\, \frac{1}{N}  \, 
\frac{1}{\lambda_{+}\! -\! \lambda_{-}}  \, 
\frac{ e^{N(A_{+} - A_{-}) }  } 
{\sqrt{ A^{\prime\prime}_{+} A^{\prime\prime}_{-}}} \, .
                                                              \label{q18a}
\end{equation} 
This leads to an oscillatory correction to the mean DOS of the following form
\begin{equation} 
\delta \nu_{osc}(E) = \frac{1}{N \pi}\,  
\frac{(-1)^{N+1}}{4 - E^2}\, 
\cos [N\left( 2\theta   + \sin 2\theta \right)]
= 
\frac{1}{N 4 \pi^3}\,  \frac{1}{\nu_0^2(E)}\, 
\cos \left[
2\pi N\int\limits_{-2}^E \nu_0(\tilde E) d \tilde E + \pi \right]
\, ,
                                                              \label{q19}
\end{equation} 
where $\sin \theta \equiv E/2$. This is indeed the correct $1/N$ oscillating 
correction, as can  be checked directly using the exact, finite $N$, expression 
for the DOS \cite{Mehta}: 
\begin{equation} 
\nu(E) = \frac{e^{-NE^2/2}}{\sqrt{2\pi N}\, 2^N\, \Gamma(N)} 
\left[ 
H_{N}\left(E\sqrt{{N\over 2}} \right) H_{N}\left(E\sqrt{{N\over 2}} \right)- 
H_{N-1}\left(E\sqrt{{N\over 2}} \right)
H_{N+1}\left(E\sqrt{{N\over 2}} \right) \right] 
                                                              \label{q20}
\end{equation} 
and employing the following  integral representation of the  Hermite 
polynomials: 
\begin{equation} 
H_{N+k}\left(E\sqrt{{N\over 2}} \right) = 
\frac{(-2i)^{N+k} }{\sqrt{\pi} }
\left(\sqrt{{N\over 2}} \right)^{N+k+1} 
\int\limits_{-\infty}^{\infty} d\lambda\, e^{-NA(\lambda,E)}\, \lambda^k \, .
                                                              \label{q21}
\end{equation} 
Evaluating this integral in the saddle point approximation, one obtains 
the ``semi-circle'' as the leading term and Eq.~(\ref{q19}) as the $1/N$ 
correction. We do not discuss here the $1/N$ correction to the smooth 
part of the DOS, which may be easily evaluated by expanding near 
the trivial $p=0$ saddle point \cite{Zirnbauer84,Itoi}. 

Notice that in deriving this result all $n+1$ saddle point 
manifolds in the $\hat Q$ space 
were taken into account. However, only two of them appear to contribute in the 
$n\to 0$ limit. In the leading order in   $N\to \infty$ only the 
trivial 
one $\hat Q\sim \openone$ remains, whereas the $p=1$ is responsible for the 
oscillatory $1/N$ contribution to the DOS. We shall see below that in the case 
of the two--point correlation function non--trivial saddle point manifolds 
contribute already to the leading order.

These corrections due to the replica symmetry braking  saddle point,
which lead to an oscillatory behaviour of the density of states
in the range $\lambda \in [-2,2]$, also lead to an exponentially
small (in $N$) tail of the density of states outside of the interval
$[-2,2]$. This fact  was first noticed by Cavagna, Giardina and Parisi
\cite{cav}, who also showed that the $p=1$ saddle point reproduces
the correct exponentially small tail.

\section{Two--Point Correlation Function}
\label{s3}

The two--point correlation function is defined as
\begin{equation} 
S_2(E, E') = N^{-2}\,
\overline{\Tr(E-H)^{-1} \Tr(E'-H)^{-1} }=
N^{-2}\, \overline{{\partial \ln Z(E) \over \partial E}
{\partial\ln Z(E')\over \partial E'}} 
\, ,
                                                              \label{w1}
\end{equation}  
where complex energies $E$ and $E'$ have negative and positive imaginary parts 
correspondingly.  
Introducing two sets of replicas with   sizes $n$ and $n'$ respectively 
to handle each of the logarithms in this expression, one obtains 
\begin{equation} 
S_2(E,E') = \lim_{n,n'\to 0} \  {1\over n n'} S_2^{(n+n')}(E,E')\, ; \hskip 1cm
S_2^{(n+n')} 
= { 1 \over N^{2}} {\partial ^2 \over \partial {E} \partial E'}
 \overline{Z(E)^n Z(E')^{n'}}\, .  
                                                             \label{w2}
\end{equation}
We  introduce again, for the sake of regularization, 
the function $Z^{(n+n')} (\hat E)$,
where $\hat E$ now, and in the rest of this section, is a diagonal 
$(n+n')\times (n+n')$ matrix of  the form 
$\hat E =\mbox{diag}\{E_1,\ldots E_{n},E_{n+1},\ldots E_{n+n'} \}$; the limit 
$E_1,\ldots E_{n}\to E$;  $E_{n+1},\ldots E_{n+n'} \to E'$ will be taken at 
the 
appropriate stage. The next steps are exactly identical to those of the previous 
section, up to the change of
$n$ into $n+n'$: averaging over $H$, decoupling with the $(n+n')\times (n+n')$  
matrix field $\hat Q$  and integrating over the group $U(n+n')$ using 
the Itzykson--Zuber integral. This leads to the
following result for the function  $Z^{(n+n')} (\hat E)$ 
(equivalent to that of Eq.~(\ref{q9})):
\begin{equation} 
\overline{Z^{(n+n')}(\hat E) } =  
c_{n+n'}(-i)^{N (n+n')}  
\left( \frac{\pi}{N} \right)^{(n+n')(n+n'-1)/2 } 
\int\!\!  d[\hat \Lambda] \, 
{\Delta_{n+n'}(\hat \Lambda) \over \Delta_{n+n'}(i\hat E) } 
(\det \hat \Lambda)^N 
 \exp \left\{ 
-{N\over 2} \sum\limits_{j=1}^{n+n'} (\lambda_j  - iE_j)^2 \right\}\, ,
                                                              \label{w3}
\end{equation}  
where  $\hat \Lambda$ is a diagonal $(n+n')\times (n+n')$  
matrix containing the eigenvalues of  the $\hat Q$ matrix.
Once again, since the integral is a totally antisymmetric function of 
$E_i$, there are no poles at $E_i=E_j$. The next step is to take the limit 
$E_j\to E$ for $j=1\ldots n$ and  $E_{j'}\to E'$ for $j'=n+1\ldots n+n'$. 
The corresponding limit is calculated in the appendix \ref{app1},
the result is: 
\begin{equation} 
\overline{Z^n(E) Z^{n'}(E')  } = c_n' c_{n'}' 
\int \!\! d[\hat \Lambda] 
\Delta_{n}^2 (\hat \Lambda^{(n)}) \Delta_{n'}^2 (\hat \Lambda^{(n')})
\frac{ \prod\limits_{j=1}^{n} \prod\limits_{j'=n+1}^{n+n'} \!\! 
(\lambda_{j'} - \lambda_j) }{[i(E'-E)]^{nn'} }  
\exp\! \left\{ \!\! 
-\!N\! \sum\limits_{j=1}^{n} A(\lambda_j,E) 
\! -\!N\!\!\!\!\! 
\sum\limits_{j'=n+1}^{n+n'}\!\!\! A(\lambda_{j'},E')\! \right\}\, ,
                                                              \label{w5}
\end{equation}  
where $\hat \Lambda^{(n)}\equiv \mbox{diag}\{\lambda_1\ldots \lambda_n\}$
and $\hat \Lambda^{(n')}\equiv \mbox{diag}\{\lambda_{n+1}\ldots 
\lambda_{n+n'}\}$.
Differentiation with respect to $E$ and $E'$ gives the replicated 
two--point correlation function. Non--trivial correlations exist only in the
range where $E-E'$ is of the order of the level spacing, namely $1/N$. 
We introduce thus  scaling variables $\eps=N E$ and $\eps'=N E'$ and 
restrict ourselves to the vicinity of the center of the band $|\eps|,\ |\eps'|
\ll N$. 
The large $N$ limit of the correlation function 
at fixed $\eps,\eps'$ is given by: 
\begin{eqnarray} 
S_2^{(n+n')}(\eps,\eps')  = c_n' c_{n'}' N^{nn'} \!\! 
\int \!\! d[\hat \Lambda] &&
\Delta_{n}^2 (\hat \Lambda^{(n)}) \Delta_{n'}^2 (\hat \Lambda^{(n')})
\frac{ \prod\limits_{j=1}^{n} \prod\limits_{j'=n+1}^{n+n'} 
(\lambda_{j'} - \lambda_j) }{[i(\eps'-\eps)]^{nn'} }
\exp\! \left\{ \!\! 
-\!N\! \sum\limits_{j=1}^{n} A \left(\lambda_j,{\eps\over N} \right)  
\! -\!N\!\!\!\!\! 
\sum\limits_{j'=n+1}^{n+n'}\!\!\! 
A \left(\lambda_{j'},{\eps'\over N} \right)\!  \right\} 
                                      \nonumber     \\ 
&&\times \left[ -\sum\limits_{j=1}^{n} \lambda_j \!\!\! 
\sum\limits_{j'=n+1}^{n+n'}  \!\!\!  \lambda_{j'}
-\frac{inn'}{\eps'-\eps} \left(
\sum\limits_{j=1}^{n} \lambda_j - \sum\limits_{j'=n+1}^{n+n'} \lambda_{j'}
\right) -\frac{nn'(1+nn')}{(\eps'-\eps)^2} 
\right]
\, . 
                                                              \label{w6}
\end{eqnarray} 
In the large $N$ limit the correlation function may be evaluated using
the saddle point approximation. 
The saddle points of the integral in Eq.~(\ref{w6}) are 
given by $\lambda_j = \pm 1$, for $j=1,\dots n+n'$, which is the zero energy
limit  of $\lambda_\pm(E)$ discussed in the previous section. 
Altogether there are $2^{n+n'}$ distinct saddle points. 
Each one of them may be 
parametrized in the following manner
\begin{equation} 
\hat\Lambda = \mbox{diag} \{
\underbrace{-1,\ldots -1}_{p}, 
\underbrace{+1,\ldots +1}_{n-p}, 
\underbrace{+1,\ldots +1}_{p'}, 
\underbrace{-1,\ldots -1}_{n'-p'}  \} \, ,
                                                              \label{w7a}
\end{equation} 
where $0\leq p\leq n$ and $0\leq p'\leq n'$. Given $p$ and $p'$,
 there are $C_n^p C_{n'}^{p'}$ 
such saddle points. 
As before, one must be careful with the saddle point 
calculation because of the vanishing prefactors. 
The method we shall follow is the same as in the previous section. 
One introduces  variables $\xi_j,\xi_{j'}'$, describing
fluctuations around the saddle point 
\bea
\lambda_i= &-1& +\xi_i/\sqrt{N}\,; \ \ \ i=1 \ldots p\, , \\ \nn
\lambda_j= &+1& +\xi_j/\sqrt{N}\,; \ \ \ j=p+1 \ldots n\, , \\ \nn
\lambda_{i'} = &+1& +\xi_{i'}'/\sqrt{N}\,; \ \ \ i'= n+1 \ldots n+p'\,, \\
\lambda_{j'} = &-1& +\xi_{j'}'/\sqrt{N}\,; \ \ \ j'=n+p'+1 \ldots n+n' \, ,
                                                          \label{w7b}
\eea
and groups them into diagonal matrices 
$\hat \Xi_-  \equiv  \mbox{diag}\{\xi_1\ldots \xi_p\}$,
$\hat \Xi_+  \equiv  \mbox{diag}\{\xi_{p+1}\ldots \xi_n\}$, 
$\hat \Xi_+' \equiv  \mbox{diag}\{\xi_{n+1}'\ldots \xi_{n+p'}'\}$,
$\hat \Xi_-' \equiv  \mbox{diag}\{\xi_{n+p'+1}'\ldots \xi_{n+n'}'\}$.
For large $N$   the determinants are decomposed as:
 \bea
\Delta_{n} (\hat \Lambda^{(n)})&\simeq& 
\(({1 \over \sqrt{N}}\))^{p(p-1)/2+(n-p)(n-p-1)/2}
2^{p(n-p)}
\Delta_p(\hat \Xi_-) \Delta_{n-p}(\hat \Xi_+) \, , 
 \\
\Delta_{n'} (\hat \Lambda^{(n')})&\simeq& 
\(({1 \over \sqrt{N}}\))^{p'(p'-1)/2+(n'-p')(n'-p'-1)/2}
(-2)^{p'(n'-p')}
\Delta_{p'}(\hat \Xi_+') 
\Delta_{n'-p'}(\hat \Xi_-') \, .
                                                        \label{det}
\eea
Finally the 
remaining factor in Eq.~(\ref{w6})  takes the form 
\be
\prod\limits_{j=1}^{n} \prod\limits_{j'=n+1}^{n+n'} 
(\lambda_{j'} - \lambda_j) \simeq
{
    2^{pp'}(-2)^{(n-p)(n'-p')} 
\over
   (\sqrt{N})^{p(n'-p')+(n-p)p'}
} \, 
\[[ \prod_{i=1}^p   \prod_{j'=n+p'+1}^{n+n'} (\xi_{j'}'-\xi_{i})\]]
 \times 
\[[ \prod_{j=p+1}^n \prod_{i'=n+1}^{n+p'}    (\xi_{i'}'-\xi_{j})\]]
\,  .
                                                      \label{w7_prelim}
\ee 

Expanding the exponent in Eq.~(\ref{w6}) around the saddle point, 
Eq.~(\ref{w7a}), 
\be
NA\left(\lambda_j,{\eps\over N} \right) \simeq {N\over 2} - N\ln(\pm 1)
\mp i\eps - {\eps^2\over 4N} + 
\left( \xi_j - {i\eps \over 2\sqrt{N}} \right)^2 +
O\left({1\over N^{3/2} }\right)\, , 
                                                        \label{A_sp}
\ee
one finds that the integrals over $\xi_j$ and $\xi_{j}'$ 
 may be expressed as
$I_{p,n'-p'}\((i(\eps-\eps') /2 \sqrt N\))\times 
I_{n-p,p'}\((i(\eps-\eps')/2\sqrt N\))$,
where the function $I_{r,s}(a)$ is
a generalization of Selberg's integral, defined as:
\be
I_{r,s}(a) \equiv 
\int d[\hat X] d[\hat Y] \, 
\Delta_r(\hat X) \Delta_s(\hat Y) \Delta_{r+s}(\hat X \oplus \hat Y) 
\exp \left\{-  \sum\limits_{j=1}^r (x_j-a)^2 - 
               \sum\limits_{k=1}^s y_k^2   \right\} \, .
                                                           \label{Selb_mod}
\ee
Here $\hat X,\hat Y$ and $\hat X \oplus \hat Y$ are diagonal matrices: 
$\hat X=\mbox{diag}\{x_1 \ldots x_r\}$,
$\hat Y=\mbox{diag}\{y_1 \ldots y_s\}$, and 
$\hat X \oplus \hat Y =\mbox{diag}\{x_1 \ldots x_r y_1 \ldots y_s \}$.
We show in  appendix \ref{app3}  that this  integral
is given by 
\be
I_{r,s}(a)= 
2^{-(r^2+s^2)/2 } \Omega_r \Omega_s (- a)^{rs}\, , 
                                                           \label{Selb_sol}
\ee
where $\Omega_r$ is  the usual Selberg integral, Eq.~(\ref{q14}).
Up to an overall constant factor which goes to one in the limit $n,n' \to 0$,
one has thus:
\be
C_{n}^{p}  C_{n'}^{p'} \  
I_{p,n'-p'}\(({i(\eps-\eps') \over2 \sqrt N}\))
I_{n-p,p'}\(({i(\eps-\eps') \over2 \sqrt N}\))= 
F_n^p F_{n'}^{p'}
\((i(\eps' - \eps) \over \sqrt N\))^{p(n'-p')+(n-p)p'} 
2^{-(p-p')^2} \ .
                                                           \label{Selb_fin}
\ee
Grouping all the terms, one  finally obtains 
\begin{eqnarray} 
S_2^{(n+n')}(\eps,\eps')  = &&
\sum\limits_{p=0}^{n}  \sum\limits_{p'=0}^{n'}
F_{n}^{p}  F_{n'}^{p'} 
\frac{(-1)^{pp'}\,  2^{-3p^2-3p'^2+4pp'} N^{(n-p+p')(n'-p'+p)} }
{[i(\eps'-\eps)]^{nn'-p(n'-p')-(n-p)p'} }   \, 
e^{2i(p'\eps' - p\eps) + i\pi N(p-p') }
\times
                                                \nonumber  \\
&& 
\left[
(n-2p)(n'-2p')
-\frac{inn'}{(\eps'-\eps)} \left(n-2p+n'-2p' \right) 
-\frac{nn'(1+nn')}{(\eps'-\eps)^2}  
\right] \, , 
                                                              \label{w8}
\end{eqnarray}
where we have omitted an inessential factor $const^{O(n)}$. 

Employing the fact that $F_{n}^{p>n} =F_{n'}^{p'>n'}=0$, one may extend 
summations 
over $p$ and $p'$ to infinity and then perform the analytical continuation, 
$n,n'\to 0$. Due to the properties of the $F$--symbol (cf. Eq.~(\ref{q17}))  
only the terms with $p=0,1$ and $p'=0,1$ contribute in the replica limit. 
We need to evaluate the contributions of these four
saddle points.
Recalling that $F_{n}^{0} = F_{n'}^{0}=1$, one finds for the 
$p=p'=0$ contribution to the
two--point correlation function, 
$S_2(\eps,\eps') = \lim_{n,n' \to 0} (n n')^{-1} S_2^{(n+n')}$:  
\begin{equation} 
{S_2(\omega)}_{(p=p'=0)} = \left(1-\frac{1}{\omega^2} \right)\, , 
                                                              \label{w9}
\end{equation}  
where $\omega\equiv \eps - \eps'$. 
This is the result obtained in 
the  perturbation theory around the usual saddle point  
\cite{Altshuler86}. 
There is, however, an  other contribution to $S_2$ 
originating from the  saddle points with $p=p'=1$. 
It is easily computed, using  
 the fact that $F_{n}^{1} = n$ and  $F_{n'}^{1} = n'$, and is equal to:
\begin{equation} 
{S_2(\omega)}_{(p=p'=1)} = \frac{e^{-2i \omega} }{\omega^2} \, . 
                                                              \label{w10}
\end{equation}
It is easy to check that the saddle points with $p=0,\, p'=1$ and 
$p=1,\, p'=0$ lead to $1/N$ oscillatory correction to the 
disconnected part of the two--point correlation function, cf. Eq.~(\ref{q19}). 
Adding together the two leading terms, Eqs.~(\ref{w9}), (\ref{w10}),  one 
obtains 
the final result  for the two--point correlation function : 
\begin{equation} 
S_2(\omega) = 1- \frac{1-e^{-2i\omega} }{\omega^2} = 
1 - 2i\omega^{-2}e^{-i\omega}\sin\omega\, . 
                                                              \label{w11}
\end{equation}  
Although this is the exact result \cite{Dyson,Mehta,Efetov83,Zirnbauer85}
 for the GUE for any $\omega \ll N$ , the way 
it was derived here justifies it only for $1\ll \omega \ll N$. This is because 
of 
the terms which were omitted in the expansions (\ref{det}), (\ref{w7_prelim}). 
Although they look superficially as being of order $1/\sqrt{N}$, they
can, in fact, contribute. A careful examination shows that the generalized
Selberg's integral is just the leading large $\omega$ ($\omega\gg 1$) 
contribution. 
We have thus computed only the leading term at large $\omega$ for each saddle
point, but corrections could be incorporated systematically. 
Accidentally the obtained expression appears to be exact down to zero $\omega$. 
A similar situation was already encountered in a SUSY approach, when the soft 
modes 
integrals were calculated with the saddle point method including an additional 
non--trivial saddle point \cite{Andreev95}. In this sense our new 
saddle point with $p=p'=1$ is a close analog of the SUSY 
saddle point of Andreev and Altshuler \cite{Andreev95}.

\section{Discussion of the method}
\label{discussion}

Let us add some comments on the relation between our calculations and the saddle 
point evaluation of the sigma model, Eq~(\ref{q6}) with 
$\hat E=\mbox{diag} \{ E\openone_{n}, E'\openone_{n'} \}$. 
Looking for the saddle 
point solution in the form $\hat Q = U^{-1}\hat \Lambda U$, one finds the 
solution in the form 
\begin{equation} 
\hat Q= \left( 
\begin{array}{cc}
V^{-1}  & 0 \\
0    & V'^{-1}
\end{array} \right)
\hat \Lambda
\left( 
\begin{array}{cc}
V  & 0 \\
0        &   V'
\end{array} \right)\, ,
                                                              \label{e2}
\end{equation} 
with $\hat \Lambda$ being a diagonal matrix, obeying 
$\Lambda^2 + i\hat E \hat \Lambda - \openone =0$, and arbitrary 
$V \in U(n)$; $V' \in U(n')$.
For a diagonal matrix $\hat \Lambda$ of the structure given by Eq.~(\ref{w7a}), 
there is a set of rotations belonging to the coset space
\begin{equation} 
\frac{U(n)}{U(p) U(n-p)} \times \frac{U(n')}{U(p') U(n'-p')}\, ,
                                                              \label{e3}
\end{equation} 
which leave the action  invariant, while changing the saddle point matrix,
$\hat Q$. As a result, there is a continuous saddle point manifold, which 
contains  true zero modes  of the functional integral, Eq~(\ref{q6}).
In addition there are usual  ``soft modes'' with 
 masses of order $|\eps-\eps'| \ll N$. Notice, that there are no zero modes
 around the trivial saddle point $p=p'=0$. 
The saddle point, Eq.~(\ref{w7a}), contains $n-p+p'$ components which are 
$+1$ and $n'-p'+p$ components which are $-1$.
Therefore, out of the total $(n+n')^2$  fluctuation directions 
$(n+n')^2 - 2(n-p+p')(n'-p'+p)$ are massive with the mass $N$, 
whereas the remaining 
$2(n-p+p')(n'-p'+p)$ degrees of freedom are split between 
$[n^2-p^2-(n-p)^2] + [n'^2-p'^2-(n'-p')^2]=2(p(n-p)+p'(n'-p'))$ 
zero modes (cf. (\ref{e3})) and $2nn' -2p(n'-p') -2(n-p)p'$ soft modes with 
the mass $|\eps-\eps'| \ll N$. The integrals over the zero modes must be 
calculated exactly giving rise to the volume of the coset space  (\ref{e3}).
In the regime $1\ll |\eps-\eps'| $, the integrals over both massive and soft 
modes 
may be evaluated in the Gaussian approximation giving rise to  factors 
$N^{-1/2}$ and $|\eps-\eps'|^{-1/2}$ in the number of modes power. This is 
precisely 
the structure of Eq.~(\ref{w8}), where the factor $F_n^p F_{n'}^{p'}$ 
is proportional to the volume of the coset space (\ref{e3}). 
The advantage of our method is an easy control over combinatorial factors, 
coefficients etc., otherwise it is equivalent to the 
Gaussian evaluation of the functional integral, Eq~(\ref{q6}), similar to 
that of Ref. \cite{Andreev95} for the SUSY case. 

Our method can be easily generalized for the higher order correlation 
functions. For example, calculations of the 
three--point function $S_3(E,E',E'')$ 
with, say, $E$ having negative and $E',E''$ positive imaginary parts, 
lead to the triple sum over $p,\, p',\, p''$ analogous to that of 
Eq.~(\ref{w8}). Again only the terms with $p,p',p'' = 0,1$ contribute in 
the replica limit. One may easily check that the correct result for the 
{\em connected} GUE three--point correlation function follows from the  
$p=p'=1;\, p''=0$ and $p=p''=1;\, p'=0$ terms,  
whereas all other possible combinations, including  $p=p'=p''=1$,
appears to be small in powers of $1/N$.

\section{Conclusion and perspectives}
\label{conclusion}

There are several questions raised by our computation. An obvious one
is to have a more rigorous derivation of the analytic continuation of the 
$g(x,n)$ function at small $n$ (see appendix  \ref{app2}). 
Extending this
approach to bosonic replicas and to other random matrix ensembles are among 
other open problems. 

We have presented here what we believe is the first consistent
application of the  replica method to the RMT. Our computation reconciles the
fermionic replica result with the previous 
approaches. The point of this paper is not to challenge these
previous approaches. The results which we have derived here
have been well known for years, and in fact there exist in
the litterature much stronger
results on level spacing universality (see e.g. \cite{brezin}
and references therein). The $\sigma$ model representation
itself has proven very successfull when used with the SUSY method:
in  problems of random energy levels, the SUSY technique has 
been very well developed and has allowed to derive many
results in various problems of solid state and nuclear physics
(see Refs. \cite{Efetov98,Weidenmuller98}). 
We think that our result has two interesting aspects: the mathematical
consistency on the one hand, and the possibility to use these ideas 
for a study of disordered {\em interacting} electrons. 

The previous situation in which the replica approach was considered as ill
was not satisfactory from the mathematical point of view.
Furthermore the replica method is known 
to be highly successful in  other problems such as the statistical
mechanics of classical disordered systems 
(see Ref. \cite{MPV} for a review) and localization theory 
\cite{Gruzberg96}, and its failure in the simple problem of 
eigenvalue correlations seemed strange. In this respect we would like to
comment about the replica symmetry breaking (RSB) which we have found.
 While the $\sigma$--model formulation seems to involve a $n \times n$ matrix
order parameter, similar to the one which has been discussed for instance 
in spin glass problems, the symmetry groups are very different. In spin glasses
the symmetry group of the replicated system is just the permutation group, while 
in the
$\sigma$--model there ia a larger symmetry group: in our case some version of 
the 
unitary group (depending on the type of correlation one computes, and whether 
one
would use commuting or anti-commuting replicas). Integrating over the angular 
variables 
has left us with an order parameter (the set of $n$ eigenvalues of the $\hat Q$ 
matrix) which is a vector in the replica space. Therefore the pattern of the 
replica 
symmetry
breaking which we have found is much more reminiscent of the `vector RSB', 
discussed in the study of random-field-like problems at low temperature
\cite{dotsenko}, rather than the hierarchical RSB scheme, which describes the
spin glass mean--field theory. The vector RSB may  be traced back 
to the existence of several distinct ground state configurations in a problem. 
In all  cases studied so far, there is  an infinity
of RSB saddle points, which contribute to the partition function. At the same 
time the SUSY approach cannot address these problems, 
because it is unable to estimate a sum over ground states,
but  rather computes a topological invariant, given
by the  sum over all saddle points, weighted by the parity of the 
number of unstable directions \cite{parisi83}. In the RMT the 
situation is
much simpler: SUSY is exact, and there is only a finite number of saddle
points contributing in the vector RSB (two saddle points in the DOS
computation). These facts are certainly 
related, and it would be highly desirable to understand better their connection.

The SUSY method relies crucially on the fact that the original
action (as in Eq.~(\ref{q3})) is quadratic in the field variables. In the
application to electronic system, it is thus restricted to non--interacting
electrons. The replica method does not have such a limitation, and
it is therefore capable to address problems of interacting electrons 
\cite{Finkelstein83}. It would be very interesting to see whether the new saddle 
points which
we have found have some implications in the theory of interacting electrons 
in disordered media.

\section*{Acknowledgments}
\label{s5}

We benefited a lot from the numerous discussions with A. Andreev, 
I. Gruzberg and S. Nishigaki. 
We are grateful to G. Parisi, C. Tracy and M. Zirnbauer for correspondence 
and bringing useful references to our account.  
This research was supported in part by the
National Science Foundation under grant No. PHY94-07194. A.K. was supported by 
NSF grant  DMR 93--0801. 
We thank G. Parisi for bringing to our attention the ref \cite{cav}.
\appendix
\section{}
\label{app1}

We first evaluate the integral given in Eq.~(\ref{q9}), 
in the limit $E_j \to E$, $j\in \{1\ldots n\}$. 
Denoting $E_j=E+\eta_j$,  one rewrites the integral in the following form 
\be
\zeta(E_1,\ldots,E_n) = \int d[\hat \Lambda] 
\Delta_n(\hat \Lambda)
\exp\left\{ 
-N  \sum\limits_{j=1}^{n} A(\lambda_j,E) 
+N \sum\limits_{j=1}^{n} \lambda_j (i \eta_j)
- {N \over 2} \sum\limits_{j=1}^{n} \eta_j^2 \right\} \, , 
                                                            \label{A1}
\ee
where $A(\lambda_j,E)$ is defined by Eq.~(\ref{q11a}). 
Expanding the term $\exp\{N \sum_j \lambda_j (i \eta_j)\}$ in series, one 
obtains:
\be
\zeta(E_1,\ldots,E_n) = \exp\left\{- {N \over 2} \sum_j \eta_j^2 \right\} 
\sum_{k_1,\ldots ,k_n =1}^\infty {N^{k_1+\ldots k_n} \over 
k_1! \ldots k_n!} (i\eta_1)^{k_1} \ldots (i \eta_n)^{k_n} T_{k_1 \ldots k_n} \ ,
                                                        \label{App_series}
\ee
where the tensor $T$ is a function of $E$ defined as:
\be
T_{k_1 \ldots k_n} \equiv  \int d[\hat \Lambda] 
\Delta_n(\hat \Lambda)
{\lambda_1^{k_1}} \ldots {\lambda_n^{k_n}} 
\exp\left\{ 
-N  \sum\limits_{j=1}^{n} A(\lambda_j,E)  \right\}  \, .
                                                               \label{A3}
\ee
Since the Vandermonde determinant is antisymmetric in $\lambda_j$, the tensor 
$T$
is fully antisymmetric: for any permutation $\pi$ of the $n$ indices, one has
\be
T_{k_1 \ldots k_n}=T_{k_{\pi(1)} \ldots k_{\pi(n)}} S_\pi\ ,
                                                             \label{A4}
\ee
where $S_\pi = \pm 1$ is the signature of the permutation. In particular, one 
notices that 
$T_{k_1 \ldots k_n}$ vanishes whenever two exponents $k_i$ and $k_j$ (with $i \ne 
j$) are
equal.
We are interested in the leading behavior of $\zeta$ when all $\eta_j$ go to 
zero
simultaneously. From the expression (\ref{App_series}) and the antisymmetry of 
$T$,
it is clear that the leading term is of order $\eta^{n(n-1)/2}$ and is obtained
whenever $k_1=0,k_2=1,\ldots , k_n=n-1$, or any permutation of the 
integers $0,1,\ldots n-1$, and the exponential prefactor may be neglected. 
The leading order in $\eta$ may be written as a sum over all  permutations $\pi$ 
of 
the ensemble $\{0,\ldots ,n-1 \}$:
\be
\zeta(E_1,\ldots,E_n) \simeq 
T_{01 \ldots n-1} \ N^{n(n-1)/2} {1 \over \prod_{j=0}^{n-1} j!}
\sum_\pi S_\pi (i \eta_1)^{\pi(0)} \ldots (i \eta_n)^{\pi(n-1)}
                                                            \label{A5}
\ee
In the sum over permutations one recognizes the Vandermonde determinant of the 
$i\eta_j$,
which is equal to $\Delta_n(i \hat E)$ and thus cancels the corresponding factor 
in 
the denominator of Eq.~(\ref{q9}). As for the value of $T_{01 \ldots n-1}$, 
it may be rewritten, using again the antisymmetry of $\Delta_n(\hat \Lambda)$, 
as:
\bea
T_{01 \ldots n-1} &=&
{1 \over n!} \sum_\pi S_\pi \int  d[\hat \Lambda] 
\Delta_n(\hat \Lambda)
\lambda_1^{\pi(0)} \ldots \lambda_n^{\pi(n-1)}
\exp\left\{ 
-N  \sum\limits_{j=1}^{n} A(\lambda_j,E)  \right\}  
                                   \nonumber \\
&=& {1 \over n!} \int d[\hat \Lambda] \Delta_n^2(\hat \Lambda)
\exp\left\{ 
-N  \sum\limits_{j=1}^{n} A(\lambda_j,E)  \right\} \, .
                                                            \label{A6}
\eea 
Therefore, in the limit $\eta_j \to 0$, $\zeta$ behaves as:
\be
\zeta(E_1,\ldots,E_n) \simeq   N^{n(n-1)/2} {1 \over \prod_{j=0}^{n} j!}
\Delta_n(i \hat E) \int d[\hat \Lambda] \Delta_n^2(\hat \Lambda)
\exp\left\{ 
-N  \sum\limits_{j=1}^{n} A(\lambda_j,E)  \right\} \, .
                                                           \label{A7}
\ee
This establishes expression (\ref{q10a}).

We now evaluate the $n+n'$ dimensional integral appearing in Eq.~(\ref{w3}),
in the limit where $E_j\to E$ for $j=1\ldots n$ and  $E_{j'}\to E'$ for 
$j'=n+1\ldots n+n'$.
The procedure is exactly the same as was explained  above for the one--point 
function.
We shall sketch it briefly.
One  writes $E_j=E+\eta_j$ and $E_{j'}=E'+\eta_{j'-n}'$ in terms of which the
integral reads: 
\bea
\zeta(E_1,\ldots,E_{n+n'}) = \int d[\hat \Lambda] \Delta_{n+n'}(\hat \Lambda)
&\exp&\left\{ 
-N  \sum\limits_{j=1}^{n} A(\lambda_j,E)  
-N  \sum\limits_{j'=n+1}^{n+n'} A(\lambda_{j'},E')  \right\} \times 
                                                 \nonumber \\
&\exp&\left\{ 
N \sum_{j=1}^{n}  \lambda_j (i \eta_j)
+N \sum_{j'=1}^{n'}  \lambda_{n+j'} (i \eta_{j'}')
- {N \over 2} \sum_{j=1}^{n}  \eta_j^2 
- {N \over 2} \sum_{j'=1}^{n'}  \eta_{j'}'^2 \right\} \ .
                                                           \label{A8}
\eea
We expand  the terms $\exp\{N \sum_j \lambda_j (i \eta_j)\}$ and 
$\exp\{N \sum_{j'} \lambda_{n+j'} (i \eta_{j'}')\}$ in series, and notice 
that in the
limit where all the $\eta_j$ and $\eta_{j'}'$ go to zero simultaneously and 
independently,
the leading contribution, of order
$(\eta)^{n(n-1)/2} (\eta')^{n'(n'-1)/2}$ is obtained when  powers of $i \eta_j$
(resp. $i \eta_{j}'$)  span the ensemble $\{ 0, \ldots , n-1 \}$
(resp. $\{ 0, \ldots , n'-1 \}$). The result can be written as a sum over all
permutations $\pi$ of 
the ensemble $\{0,\ldots ,n-1 \}$, and all
permutations $\pi'$ of 
the ensemble $\{0,\ldots ,n'-1 \}$, in the following form:
\be
\zeta(E_1,\ldots,E_{n+n'}) = u \ N^{{n(n-1) \over 2}+{n'(n'-1) \over 2}} 
 \((\prod_{j=0}^{n-1} j! \prod_{j'=0}^{n'-1} j'!\))^{-1}\!\!\! 
\sum_{\pi \pi'} S_\pi S_{\pi'} (i \eta_1)^{\pi(0)} \ldots (i \eta_n)^{\pi(n-1)}
 (i \eta_1')^{\pi'(0)} \ldots (i \eta_{n'}')^{\pi'(n'-1)} \ ,
                                                              \label{A9}
\ee
where $u$ is equal to:
\be
u=\int d[\hat \Lambda] \Delta_{n+n'}(\hat \Lambda) \ 
 \lambda_1^0 \lambda_2^1 \ldots
\lambda_n^{n-1} \ 
\lambda_{n+1}^0 \lambda_{n+2}^1 \ldots
\lambda_{n+n'}^{n'-1} 
\exp\left\{ 
-N  \sum\limits_{j=1}^{n} A(\lambda_j,E)  
-N  \sum\limits_{j'=n+1}^{n+n'} A(\lambda_{j'},E')  \right\}  \ .
                                                            \label{A10}
\ee
Exactly as above, this integral may be rewritten by permuting separately the 
dummy
integration variables $\lambda_1,\ldots ,\lambda_n$ and 
$\lambda_{n+1},\ldots ,\lambda_{n+n'}$. One thus obtains 
in the limit where all $\eta_j, \eta_{j'}' \to 0$, in the notations
of Eq.~(\ref{w5}):
\bea
\zeta(E_1,\ldots,E_{n+n'}) \simeq 
N^{{n(n-1) \over 2}+{n'(n'-1) \over 2}}
\((\prod_{j=0}^{n} j! \prod_{j'=0}^{n'} j'!\))^{-1}
\prod_{1\le i<j \le n} (E_j-E_i)
\prod_{n+1\le i'<j' \le n+n'} (E_{j'}-E_{i'}) \times 
                                              \nonumber     \\
 \int d[\hat \Lambda]\  \Delta_{n+n'}(\hat \Lambda)
 \Delta_n(\hat \Lambda^{(n)})\Delta_{n'}(\hat \Lambda^{(n')})
\exp\left\{ 
-N  \sum\limits_{j=1}^{n} A(\lambda_j,E)  
-N  \sum\limits_{j'=n+1}^{n+n'} A(\lambda_{j'},E')  \right\}  \ .
                                                             \label{A11}
\eea
This establishes expression (\ref{w5}).

\section{}
\label{app2}

To prove  the statement that only
the terms with $p=0$ and $p=1$ contribute to the analytic continuation
of Eq.~(\ref{q15})  at small $n$, let us 
study the function $g(x,n)=\sum_{p=0}^n F_n^p x^p$. Using the Gaussian 
decomposition of 
the factors containing exponents of $p(n-p)$ and $p^2$, one may show 
that the correlation
function $S_1^{(n)}$ is deduced from the knowledge of $g(x,n)$, where $x$  is
a complex number,  with a modulus slightly smaller than one \cite{fn1}. 
Using the fact that
$F_n^{p+1}= F_n^p \ \Gamma(p+1)/\Gamma(n-p)$, one finds  that  for integer $n$ 
the function
$g(x,n)$ satisfies the following integral equation:
\be
g(x,n)=1+ x \int\limits_0^\infty\!\! d\tau\, e^{-\tau}\! 
\int\limits_{{\cal C}} {du \over 2 \pi}\, 
{e^u \over u^n}\,  g(x \tau u,n) \, ,
                                                    \label{B1}
\ee
where ${\cal C}$ is the contour in the complex $u$ plane used for definition of 
the
function $1/\Gamma(z)$: it goes around the negative real half axis, starting 
from 
$-\infty$ to 0 with a small positive imaginary part, turning around $0$ and 
getting
back to $-\infty$ with a small  negative imaginary part, 
it thus passes around the cut of the $1/u^n$ function for $n$ non--integer. 
This integral may be probably used to define the function 
$g(x,n)$ for an arbitrary $n$, 
although some further study of this statement is needed.
Here we are interested in the behavior of $g(x,n)$ at small $n$. Writing the 
first
two terms in the small $n$ expansion as $g(x,n)=g_0(x)+n g_1(x)+\ldots$, one 
finds 
that $g_0$ and $g_1$ satisfy the following equations:
\bea
g_0(x)&=&1+ x \int\limits_0^\infty 
d \tau e^{-\tau} \int\limits_{{\cal C}} {du \over 2 \pi}\, {e^u }\, g_0(x \tau 
u) 
                                              \nonumber   \\
g_1(x)&=&x \int\limits_0^\infty 
d \tau e^{-\tau} \int\limits_{{\cal C}} {du \over 2 \pi}\, 
{e^u } \left[ g_1(x \tau u) - g_0(x\tau u) \ln u \right]\, .
                                                      \label{B2}
\eea
Assuming that $g_0$ and $g_1$ are analytic in a certain domain ${\cal D}$
near the origin,  one can compute them inside this domain by 
series expansion in powers of $x$. This leads immediately to $g_0=1$ and 
$g_1=x$, 
which
gives exactly the same answer as our heuristic arguments given in the text. To 
complete
the proof one has to find out a shape of ${\cal D}$. We believe that ${\cal D}$
is the part of the complex plane restricted by the unit circle, 
but we have not been able to prove it \cite{fn2}.

\section{}
\label{app3}

In this appendix we prove that the generalized Selberg integral, 
Eq.~(\ref{Selb_mod}), is given by the expression (\ref{Selb_sol}).
The  proof consist of the two steps:

(i) We shall prove that the series expansion of the integral in powers of 
$a$ 
starts as $a^{m}$ with $m \geq rs$. To this end we rewrite the integral as 
\be
I_{r,s}(a) = \int d[\hat X] d[\hat Y] \
\Delta_r(\hat X) \Delta_s(\hat Y) \Delta_{r+s}(\hat X \oplus \hat Y) 
\exp\left\{ - \sum\limits_{j=1}^r x_j^2 - \sum\limits_{k=1}^s y_k^2 
+2a\sum\limits_{j=1}^r x_j  - r a^2 \right\} 
                                                    \label{C1}
\ee
and notice that the integrand is equal to $\Delta_r(\hat X)$ times a totally 
antisymmetric function
of the $x_j$. This allows one to substitute in the integrand $\Delta_r(\hat X)$ 
by 
$r! \ x_1^0 x_2^1 \ldots x_r^{r-1}$. A similar observation for the $y$ variables
allows one to substitute in the integrand $\Delta_s(\hat Y)$ by 
$s! \ y_1^0 y_2^1 \ldots y_s^{s-1}$, giving:
\be
I_{r,s}(a) =  r! s! e^{ -r a^2 } 
\int d[\hat X] d[\hat Y] \
\Delta_{r+s}(\hat X \oplus \hat Y) \
[x_1^0 x_2^1 \ldots x_r^{r-1} \ y_1^0 y_2^1 \ldots y_s^{s-1} ] \ 
\exp\left\{ - \sum\limits_{j=1}^r x_j^2 - \sum\limits_{k=1}^s y_k^2 + 
2a \sum\limits_{j=1}^r x_j\right\}
                                                                  \label{C2}
\ee
The integrand is the product of a term which is totally antisymmetric in all the 
$r+s$ 
integration  variables times the factor 
$ [x_1^0 x_2^1 \ldots x_r^{r-1} \
y_1^0 y_2^1 \ldots y_s^{s-1} ]\exp\{2a \sum_j x_j\}$. 
In this factor one can expand the exponential in a power series in $a$. 
Whenever there are two of the $r+s$ variables appearing  with the same power, 
the 
the integral  is zero
as can be seen by permuting these two variables. The first non-zero contribution  
appears, thus, when the power series generates a power like 
$y_1^0 y_2^1 \ldots y_s^{s-1} x_1^{\pi(s)} x_2^{\pi(s+1)} \ldots 
x_r^{\pi(s+r-1)}$,
where $\pi$ is any permutation of the integers  $s,s+1,\ldots,s+r-1$. Such terms 
appear 
when the exponential is expanded to the order $a^{rs}$. This shows that 
the series expansion of the integral in powers of $a$ starts 
at least with the order $a^{rs}$.

(ii) We demonstrate now that $I_{r,s}$ is a polynomial in $a$ of degree {\em 
less} 
or equal to $rs$. Shifting each $x_j$ to $x_j+a$ and splitting the factor 
$\Delta_{r+s}(\hat X \oplus \hat Y)$, one may rewrite the integral as 
\be
I_{r,s}(a) =\int d[\hat X] d[\hat Y] \
\Delta_r^2(\hat X) \Delta_s^2(\hat Y) 
\exp\left\{ - \sum\limits_{j=1}^r x_j^2 - \sum\limits_{k=1}^s y_k^2 \right\} 
\prod\limits_{j=1}^r \prod\limits_{k=1}^s (y_k-x_j-a)\, .
                                                         \label{Inew_shifted}
\ee
This shows that the integral is a polynomial in $a$ of a degree less or equal to 
$rs$. 
 
As a result of (i) and (ii), $I_{r,s}(a)$ must be proportional to $a^{rs}$. 
Using 
Eq.~(\ref{Inew_shifted}), 
one finds that it may be expressed as a product of two usual Selberg integrals,
Eq.~(\ref{q14}),  as:
\be
I_{r,s}(a)=(-a)^{rs} \int d[\hat X] d[\hat Y] \
\Delta_r^2(\hat X) \Delta_s^2(\hat Y) 
\exp\left\{ - \sum\limits_{j=1}^r x_j^2 - \sum\limits_{k=1}^s y_k^2 \right\}
=(-a)^{rs} 2^{-(r^2+s^2)/2} \Omega_r \Omega_s\, .
                                                           \label{C4}
\ee


\begin{thebibliography}{99}
\bibitem{fn_sadpo}
In the limit where the imaginary part of $E$ vanishes, the actions of the 
various saddle
points differ only by phases. If one is interested only in the DOS to leading 
order at
large $N$, one may choose the imaginary part of the energy, $\eta$, to be such 
that $\eta N \gg 1$.
This makes the trivial saddle point with $p=0$ dominant over all others.
However this procedure, which smoothes the positions of individual levels, 
is not allowed if one computes oscillatory parts of correlation functions.
\bibitem{foot0}To preserve the Hermiticity one may choose the elements of the 
$\hat E$ matrix on the negative imaginary axis and perform the analytical 
continuation to the real axis after the group integration. Since the resulting 
expression is an analytic function of all $E_i$ in the lower half plane,
  this procedure is 
legitimate.
\bibitem{fn1}
If one neglects an imaginary part of $E$, 
$|x|=1$. A small negative imaginary part of $E$ implies $|x|<1$. 
This is the reason for our choice of the 
parameterization of the saddle point, where there are $p$ eigenvalues 
$\lambda_-$,
rather than the opposite choice where there would  be $p$ eigenvalues 
$\lambda_+$.
\bibitem{fn2}
One argument in favor of this is the fact that $g(x,n)$, for integer $n$, 
possesses the
symmetry $g(x,n)=x^n g(1/x,n)$; another indication comes from the fact that 
$|g(x,n)|$ is
bounded at large $n$ for a fixed complex $x$ whenever $|x|<1$, while it diverges
for $|x|>1$.
\end{thebibliography}
\end{document}